\documentclass[aps,twocolumn,english,superscriptaddress,citeautoscript,preprintnumbers,amsmath,amssymb,floatfix,footinbib]{revtex4-2}
\usepackage[breaklinks=true,colorlinks,citecolor=blue,linkcolor=blue,urlcolor=blue]{hyperref}
\usepackage{amsmath}
\usepackage{graphicx}
\usepackage{dcolumn}
\usepackage{float}
\usepackage[utf8]{inputenc}
\usepackage{color}
\usepackage[super]{nth}
\DeclareUnicodeCharacter{2212}{-}

\usepackage{mathtools}
\DeclarePairedDelimiter\bra{\langle}{\rvert}
\DeclarePairedDelimiter\ket{\lvert}{\rangle}
\begin{document}
	
\title{The magnetic, thermodynamic, and magnetotransport properties of CeGaGe and PrGaGe single crystals}
	\author{Daloo Ram}
\affiliation{Department of Physics, Indian Institute of Technology, Kanpur 208016, India}
\author{Sudip Malick}
\affiliation{Department of Physics, Indian Institute of Technology, Kanpur 208016, India}
\author{Zakir Hossain}
\email{zakir@iitk.ac.in}
\affiliation{Department of Physics, Indian Institute of Technology, Kanpur 208016, India}
\author{Dariusz Kaczorowski}
\email{d.kaczorowski@intibs.pl}
\affiliation{\mbox{Institute of Low Temperature and Structure Research, Polish Academy of Sciences,ulica Ok\'olna 2, 50-422 Wroclaw, Poland}}

\begin{abstract}

We investigate the physical properties of high-quality single crystals CeGaGe and PrGaGe
using magnetization, heat capacity, and magnetotransport measurements. Gallium-indium binary flux was used to grow these single crystals that crystallize in a body-centered tetragonal structure. Magnetic susceptibility data reveal a magnetic phase transition around 6.0 and 19.4 K in CeGaGe and PrGaGe, respectively, which is further confirmed by heat capacity and electrical resistivity data. A number of additional anomalies have been observed below the ordering temperature in the magnetic susceptibility data, indicating a complex magnetic structure. The magnetic measurements also reveal a strong magnetocrystalline anisotropy in both compounds. Our detailed analysis of the crystalline electric field (CEF) effect as observed in magnetic susceptibility and heat capacity data suggests that \textit{J} = 5/2 multiplet of CeGaGe splits into three doublets, while \textit{J} = 4 degenerate ground state of PrGaGe splits into five singlets and two doublets. The estimated energy levels from the CEF analysis are consistent with the magnetic entropy.

\end{abstract}

\maketitle
\section{Introduction}	

The competing interactions resulting from the hybridization of 4\textit{f} and conduction electrons in the strongly correlated electron system yields remarkable physical features such as the Kondo effect, heavy fermion behavior, intermediate valency, crystalline electric field (CEF) effect, magnetic ordering, superconductivity, and so on \cite{CePd2Al8, Ce3PtIn11, CeCu2Si2,CeAgAs2,PrRh2Si2,HFSC,Ce3NiSi3,Ce3Rh4Sn7}. Therefore, rare-earth-based intermetallic compounds have always been appealing because they provide a platform for investigating such exotic physical properties. The introduction of strong correlation into a nontrivial topological system results in more intriguing ground states. For instance, Ce$_3$Bi$_4$Pd$_3$ is reported to be a Weyl-Kondo heavy-fermion semimetal \cite{Ce3Bi4Pd3,Ce3Bi4(PtPd)3}, whereas Ce$_3$Bi$_4$Pt$_3$ displays a topological Kondo insulating state \cite{Ce3Bi4(PtPd)3,Ce3Bi4Pt3}. On the other hand, the coexistence of relativistic fermions and magnetism in these compounds manifests anomalous transport \cite{GdPtBi,TbPtBi,SmB6_2013, EuAs3}.

The ternary tetragonal compounds RAlX (R = La$-$Nd and Sm; X = Si and Ge) have gained attention recently since this class of compounds displays a Weyl semimetal (WSM) state \cite{LaAlSi,LaAlGe,SmAlSi,NdAlGe,CeAlGe2020,PrAlGe2020,CeAlGe2018, CeAlSi2021,PrAlGe2019,PrAlSi,NdAlSi2022}. Interestingly, RAlX crystallizes in two space groups: LaPtSi-type noncentrosymmetric space group \textit{I}4$_1$\textit{md} (No. 109) and $\alpha$-ThSi$_2$-type centrosymmetric space group \textit{I}4$_1$/\textit{amd} (No. 141). LaAlGe crystallizes in a noncentrosymmetric structure and exhibits a type-II WSM state as determined by angle-resolved photoemission spectroscopy \cite{LaAlGe}. Moreover, magnetic compounds are more intriguing because they provide a framework for investigating the interaction between relativistic Weyl fermions and magnetism. For instance, CeAlSi is a noncollinear ferromagnetic (FM) WSM that exhibits an anisotropic anomalous Hall effect \cite{CeAlSi2021} and PrAlSi is a centrosymmetric ferromagnet manifesting an anomalous Hall effect for magnetic field applied along $c$ axis \cite{PrAlSi}. In addition, the substitution of Ge with Si in PrAlGe results in a crossover from intrinsic to extrinsic anomalous Hall conductivity \cite{PrAlGe1-xSix}. Further, this family of compounds also exhibits strong magnetocrystalline anisotropy and the CEF effect \cite{CeAlGe2018,CeAlSi2021,PrAlGe2019,PrAlSi,NdAlSi2022,NdAlGe2023}. Moreover, a number of recent reports reveal the presence of a WSM state exhibiting complex magnetic structure and anomalous transport in CeAlSi \cite{CeAlSi2021}, CeAlGe \cite{CeAlGe2023}, PrAlSi \cite{PrAlSi}, NdAlSi \cite{NdAlSi2021}, and NdAlGe \cite{NdAlGe2023,NdAlGearxiv2023}.

In this report, we present a thorough investigation of the physical properties of CeGaGe and PrGaGe single crystals. CeGaGe is a FM Kondo lattice system with an ordering temperature of 5.5 K as determined in polycrystalline samples \cite{CeGaGe1993,CeGaGe1996}. However, to the best of our knowledge there is no report on PrGaGe. Here, we have grown single crystals of RGaGe (R = Ce and Pr) as well as its nonmagnetic counterpart, LaGaGe, using the flux method. Powder x-ray diffraction (XRD) patterns of crushed single crystals revealed that RGaGe (R = La$-$Pr) crystallizes in a noncentrosymmetric structure with space group \textit{I}4$_1$\textit{md}. Magnetic susceptibility data along with the heat capacity and electrical resistivity data suggest a complex magnetic ordering and CEF effect in \mbox{CeGaGe} and PrGaGe. Further, both compounds exhibits negative magnetoresistance near the magnetic ordering temperature.

\section{Experimental details}

\begin{figure*}
	\centering
	\includegraphics[width=17cm, keepaspectratio]{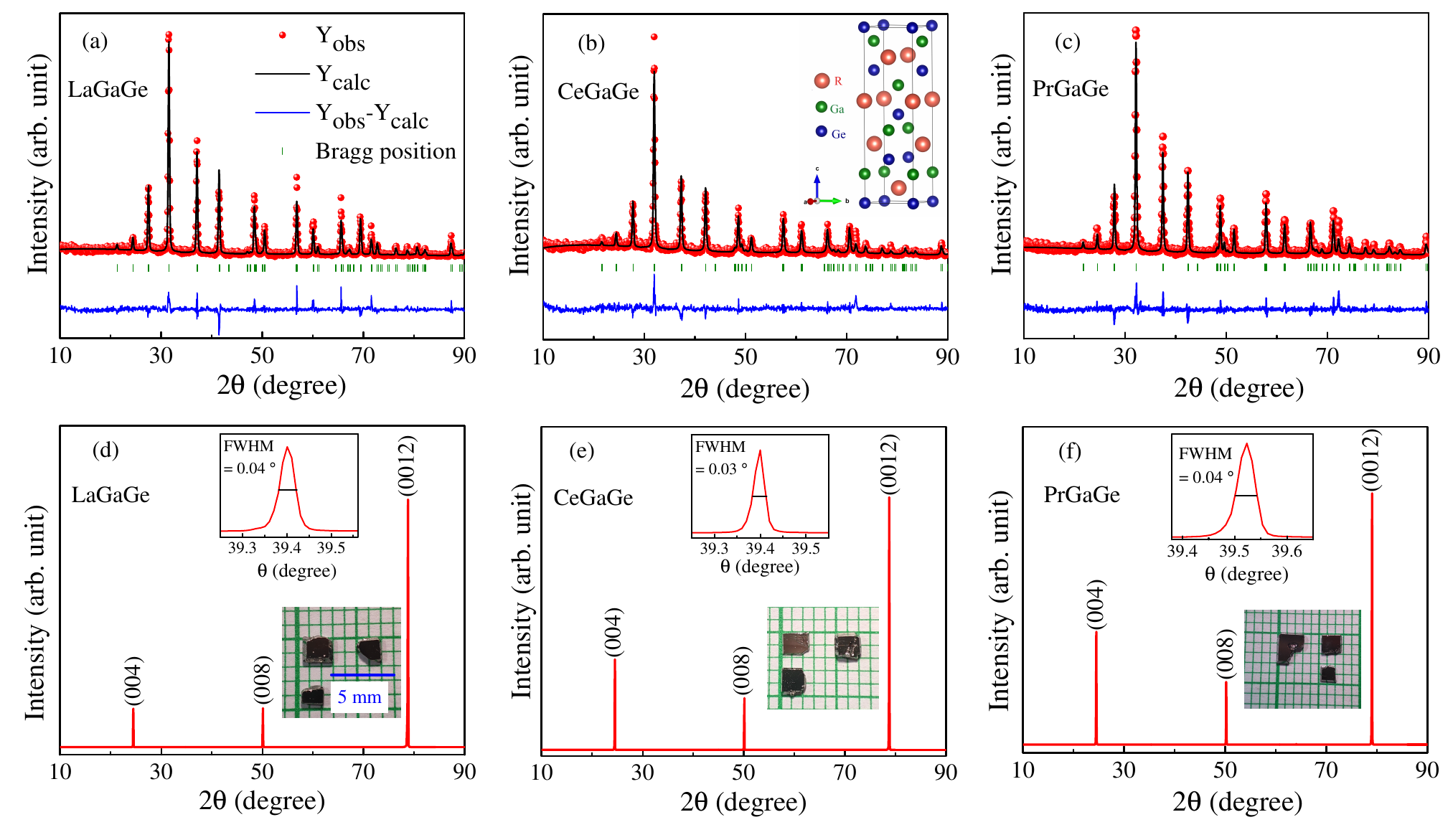}
	\caption{\label{XRD}Powder XRD patterns along with Rietveld refinement of crushed single crystals (a) LaGaGe, (b) CeGaGe, and (c) PrGaGe recorded at room temperature. The solid black line and red circles represent the calculated pattern and experimental data. The blue line shows difference between experimental and calculated intensities. The olive bars mark the Bragg positions. Inset of (b) displays the crystal structure of RGaGe. Single crystal XRD patterns are presented (d) LaGaGe, (e) CeGaGe, and (f) PrGaGe. Insets of (d), (e), and (f) show optical images of single crystals (lower) and the rocking curves (upper).}
\end{figure*}

High-quality single crystals of RGaGe (R = La$-$Pr) were grown using gallium-indium binary flux. Constituent elements La (99.9\%, Alfa Aesar), Ce (99.9\%, Alfa Aesar), Pr (99.9\%, Alfa Aesar), Ga (99.9999\%, Alfa Aesar), Ge (99.999\%, Alfa Aesar), and In (99.99\%, Alfa Aesar) are taken in the molar ratio R:Ga:Ge:In = 1:2:1:8. The elements were mixed up properly and placed in an alumina crucible. Next, the crucible was sealed inside quartz tube with partial argon pressure. The sealed quartz tube was put into a muffle furnace and heated to 1050 $^\circ$C for 24 h. Then the furnace was slowly cooled down to 500 $^\circ$C at a rate of 3 $^\circ$C/h. At this point the excess flux was removed by centrifuging. Platelike shiny single crystals with a typical dimension of 3 $\times$ 3 $\times$ 0.4 mm$^3$ were obtained as shown in lower insets of bottom panel of Fig \ref{XRD}.

The structural characterization of obtained crystals was carried out using XRD method in a PANalytical X’Pert PRO diffractometer with Cu K$_{\alpha1}$ radiation. The chemical compositions of the crystals were checked by energy-dispersive spectroscopy (EDS) in a JEOL JSM-6010LA electron microscope. Electrical resistivity and magnetotransport measurements were performed in a Quantum Design physical property measurement system (PPMS) utilizing the usual four-probe method. Heat capacity measurements were performed using the relaxation method in the heat capacity option of PPMS. The magnetic properties were measured using a Quantum Design magnetic property measurement system.

\section{Results and discussion}

\begin{table}
	\label{Lattice}
	\caption{The obtained lattice parameters of RGaGe from powder XRD data recorded at room temperature.} 
	
	\centering
	\addtolength{\tabcolsep}{1pt}
	\begin{tabular}{cccc}
		\hline\hline \\[0.01ex]
		
		Lattice parameters	~~& LaGaGe~~~~ & CeGaGe~~~~ & PrGaGe~~~~ \\[1.5ex]
		\hline  \\[0.01ex]
		
		$a$ (\AA)  & 4.3542(2)          & 4.2911(5)          & 4.2599(4)        \\[1.5ex]
		$b$ (\AA)  & 4.3542(2)          & 4.2911(5)          & 4.2599(4)        \\[1.5ex]
		$c$ (\AA)  & 14.581(1)          & 14.576(2)          & 14.521(2)        \\[1.5ex]
		$V$ (\AA$^3$) & 276.43(3)       &268.39(6)            & 263.51(5)        \\[1.5ex]
		\hline
		\hline
	\end{tabular}
\end{table}
\subsection{Crystal structure}

Powder XRD patterns of crushed crystals, recorded at room temperature, are shown in Figs. \ref{XRD}(a)$-$\ref{XRD}(c). The analysis of the powder XRD patterns using the Rietveld refinement method suggest all three RGaGe (R = La$−$Pr) compounds crystallize in a body-centered tetragonal structure with space group \textit{I}4$_1$\textit{md} (No. 109) \cite{RAlGe2019}. The estimated lattice parameters as presented in Table \textcolor{blue}{I} agree well with the previous report \cite{CeGaGe1993}. A schematic diagram of the crystal structure is displayed in the inset of Fig. \ref{XRD}(b). The structure consists of four formula unit per unit cell, in which all three atoms (R, Ga, and Ge) have 4\textit{a} Wyckoff site. There are stacking of R, Ga, and Ge layers along (001) direction. This structure has two vertical mirror planes ($\sigma_v$) but lacks a horizontal mirror plane ($\sigma_h$), resulting in inversion symmetry breaking. The observed single crystal XRD pattern of RGaGe can be indexed by (00\textit{l}) Miller indices, as depicted in Figs. \ref{XRD}(d)$-$\ref{XRD}(f), indicating that the crystallographic $c$ axis is perpendicular to the flat plane of single crystals. Moreover, the peaks are extremely sharp, suggesting the high quality of the single crystals. Further, the full width at half maximum (FWHM) of the rocking curve [see upper insets of Figs. \ref{XRD}(d)$-$\ref{XRD}(f)] for the peak (0012) is about 0.04$^\circ$, which is very small, confirming that all single crystals are of excellent quality. EDS data collected from multiple points and areas on the surface of crystals reveal expected stoichiometry.

\subsection{Magnetic properties}

\begin{figure*}
	\centering
	\includegraphics[width=17cm, keepaspectratio]{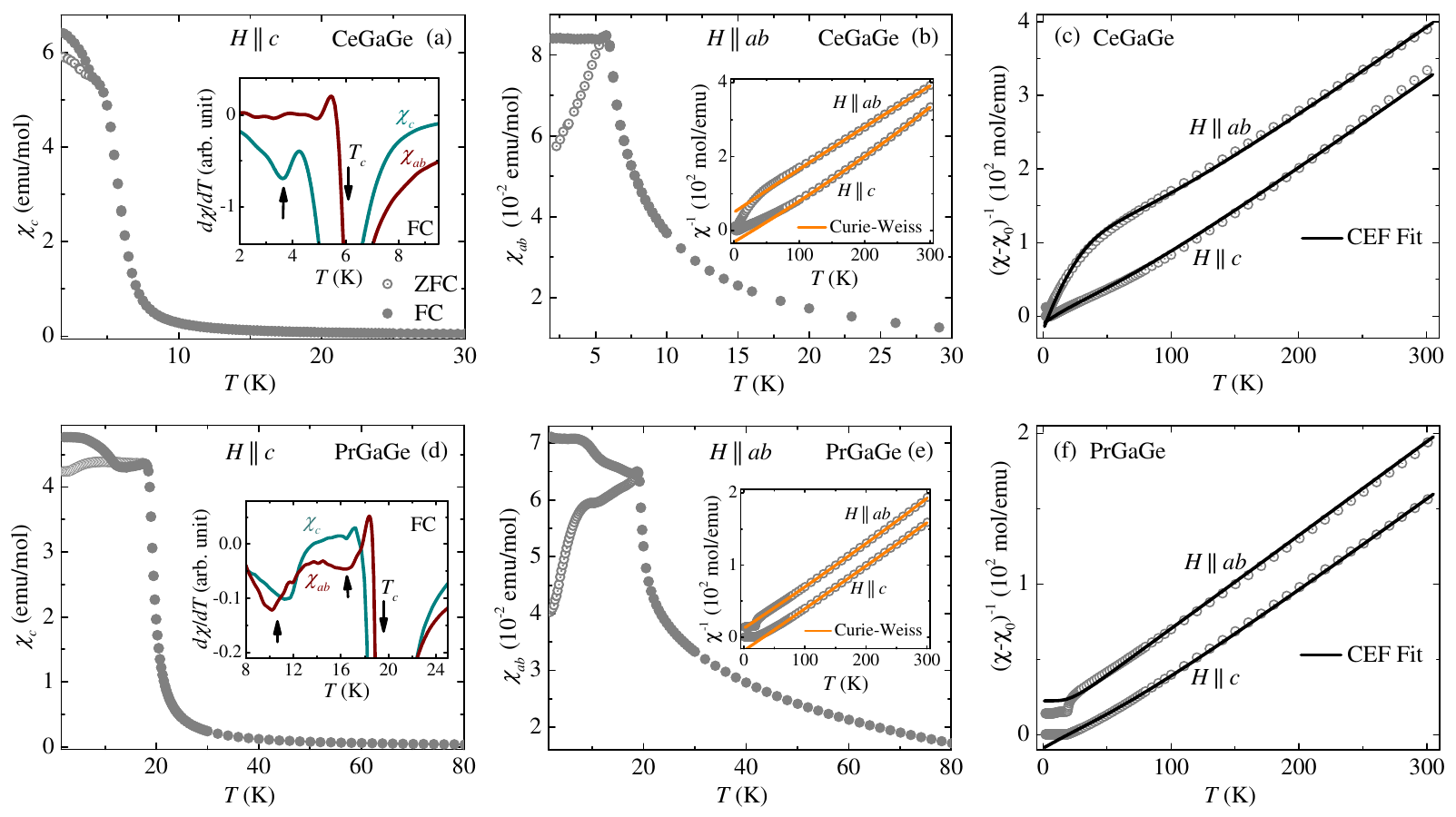}
	\caption{The temperature-dependent magnetic susceptibility ($\chi$) measured at 0.1 T along (a) $H \parallel c$ and (b) $H \parallel ab$ of CeGaGe. The open and filled circles represent magnetic susceptibility measured in ZFC and FC configurations, respectively. The $\chi$(\textit{T}) of PrGaGe for ZFC and FC configurations was measured at 0.1 T for (d) $H \parallel c$ and (e) $H \parallel ab$, respectively. The insets (a) and (d) display the temperature-dependent d$\chi$/d\textit{T} for FC configuration for CeGaGe and PrGaGe, respectively. The insets of (b) and (e) depict the inverse magnetic susceptibilities of CeGaGe and PrGaGe at 0.1 T, respectively, as a function of temperature. The orange lines represent Curie-Weiss fitting. The temperature-dependent inverse magnetic susceptibility of (c) CeGaGe and (f) PrGaGe at field 0.1 T. The black solid lines in (c) and (f) represent the calculated inverse susceptibility using CEF model.}
	\label{MT}
\end{figure*}

The temperature-dependent magnetic susceptibilities $\chi_c$ ($H\parallel c$) and $\chi_{ab}$ ($H\parallel ab$) for zero-field cooled (ZFC) and field cooled (FC) configurations at 0.1 T for both compounds are shown in Figs. \ref{MT}(a), \ref{MT}(b), \ref{MT}(d), and \ref{MT}(e). The sudden upturn in $\chi_c$ around $T_C$ = 6.0 K in \mbox{CeGaGe} and $T_C$ = 19.4 K in PrGaGe suggest a FM phase transition. However, $\chi_{ab}$ slowly increases with decreasing temperature down to $T_C$ in both compounds. Interestingly, the ZFC and FC data exhibit a bifurcation below $T_C$ along both crystallographic directions, more prominent along $H\parallel ab$. Moreover, $\chi_{ab}$ for ZFC drops below $T_C$ in an AFM fashion. The value of $\chi_{c}/\chi_{ab}$ is about 10 above T$_C$ and reaches a value of $\sim$ 105 at 2 K in both compounds, which suggests strong magnetocrystalline anisotropy in these compounds. A similar temperature dependence with strong magnetic anisotropy was observed in same family of compounds RAlX (R = Ce$-$Nd; X = Ge and Si) \cite{CeAlSi2021,PrAlGe2019,PrAlSi,NdAlSi2022}. In the magnetically ordered state, $\chi$(\textit{T}) shows an anomaly around 3.7 K in CeGaGe, and two anomalies around 11.0 and 16.5 K in PrGaGe, as is evident from the $ d\chi/dT$ data, presented in the insets of Figs. 2(a) and 2(b). The observed anomalies may be associated with the transition from incommensurate to commensurate order, which has recently been observed in WSM NdAlSi \cite{NdAlSi2021,NdAlSi2022} and NdAlGe \cite{NdAlGe2023,NdAlGearxiv2023}. Therefore, a comprehensive investigation utilizing neutron diffraction is required to reveal the complex magnetic structure of CeGaGe and PrGaGe. The inverse magnetic susceptibility data are plotted as a function of temperature and fitted well above $T_C$ with the modified Curie-Weiss law, $\chi\left(T\right)=\chi_0 + C/(T-\Theta_P)$, where \textit{C} is the Curie constant, $\Theta_P$ is the paramagnetic Curie temperature, and $\chi_0$ is temperature-independent magnetic susceptibility. The insets of Figs. \ref{MT}(b) and \ref{MT}(e) illustrate the fitting of Curie-Weiss law for CeGaGe and PrGaGe, respectively. The obtained fitting parameters are listed in Table \textcolor{blue}{II}. The negative values of $\chi_{0}$ may appear due to the diamagnetic contribution of sample holder. The positive $\Theta_P$ along the \textit{c} axis indicates the dominant FM exchange interactions whereas negative $\Theta_P$ along the \textit{ab} plane suggests an AFM coupling. Similar values of $\Theta_P$ were also observed in isostructural compounds such as CeAlGe \cite{CeAlGe2018}, CeAlSi \cite{CeAlSi2021}, and PrAlGe \cite{PrAlGe2019,PrAlGe2020}. The calculated effective moments $\mu_{eff}$ of CeGaGe and PrGaGe are very close to Ce$^{3+}$ (2.54 $\mu_B$) and Pr$^{3+}$ (3.58 $\mu_B$) ions, respectively. Below 100 K, the inverse susceptibility deviates from the Curie-Weiss law and exhibits a hump around 50 K, as depicted in the insets of Figs. \ref{MT}(b) and \ref{MT}(e). Such a hump appears due to the CEF effect as observed in numerous rare-earth compounds such as CeAgAs$_2$ \cite{CeAgAs2}, PrSi \cite{PrSi}, and Pr$_2$Re$_3$Si$_5$ \cite{Ce2Re3Si5}.

\begin{table}
	\label{CW_fit}
	\caption{The estimated value of $\chi_0$, $\mu_{eff}$, and $\Theta_P$ from modified Curie-Weiss fit of CeGaGe and PrGaGe.} 
	\vskip .1cm
	\begin{tabular}{ccccc}
		\hline
		\hline
		&\multicolumn{2}{c}{CeGaGe}&	\multicolumn{2}{c}{PrGaGe}\\
		
		&~~\textit{H} $\parallel$ \textit{c}&~~~~\textit{H} $\parallel$ \textit{ab}~~~&~~~\textit{H} $\parallel$ \textit{c}~~~~& \textit{H} $\parallel$ \textit{ab}~~~~\\[1.5ex]
		\hline\\[0.01ex]
		
		$\chi_0$ (10$^{-4}$emu/mol)   &-3.97 &-0.16  &-1.23  &-3.71\\[1.5ex]
		$\mu_{eff}$ ($\mu_B$)         &2.71  &2.59   &3.70   & 3.78\\[1.5ex]
		$\Theta_P$ (K)                &28.7  &-41.9 &32.3 	 &-21.9\\[1.5ex]
		
		\hline
		\hline
		
	\end{tabular}
\end{table}

\begin{table}
	\label{CEF_CeGaGe}
	\centering 
	\caption {CEF parameters, energy levels and the corresponding wave functions for CeGaGe}
	\vskip .1cm
	\begin{tabular}{ccccc}
		\hline
		\hline \\[0.01ex]
		\multicolumn{5}{c}{CEF parameters}\\[1.5ex]
		\hline \\[0.01ex]
		~~$B_2^0$ (K) ~~& $B_4^0$ (K)~~& $B_4^4$ (K) ~~&~ $B_6^0$ (K) & $B_6^4$ (K)  \\[1.5ex]
		
		-6.40 $\pm$ 0.20 & -0.418 $\pm$ 0.035 & 3.32 $\pm$ 0.270 & 0 & 0  \\[1.5ex] 
		\hline\\
		   \multicolumn{2}{c}{$\lambda_z$ = 10.0 $\pm$ 1.7 (mol/emu)}~~~~~~~ & \multicolumn{3}{c}{$\lambda_{x,y}$ = 18.7 $\pm$ 1.8 (mol/emu)}
		     ~~~ \\[1.5ex]
		
		\hline
		
	\end{tabular}
	\begin{tabular}{ccccccc}
			
			\hline\\
		\multicolumn{7}{c}{Energy levels and wave functions} \\[1.5ex]
		\hline \\[0.01ex]
     ~~~~E (K) ~~~~~&~~~ $\ket{+5/2}$~&~$\ket{+3/2}$~&~$\ket{+1/2}$~&~$\ket{-1/2}$~&~$\ket{-3/2}$~&~$\ket{-5/2}$\\[1.5ex]
		\hline\\
      251.4 $\pm$ 12.6 &  -0.38     &      0     &    0     &  0   &  -0.92   &    0   \\[1.5ex]
      251.4 $\pm$ 12.6 &    0       &    -0.92   &    0     &  0   &    0     & -0.38 \\[1.5ex]	
      126.9 $\pm$ 7.4  &    0       &      0     &    1     &  0   &    0     &    0  \\[1.5ex]	
      126.9 $\pm$ 7.4  &    0       &      0     &    0     &  1   &    0     &    0  \\[1.5ex]	
      0                &    0.92    &      0     &    0     &  0   &  -0.38   &    0   \\[1.5ex]
      0                &    0       &   -0.38    &    0     &  0   &    0     &   0.92  \\[1.5ex]
		
		\hline
		\hline
		
	\end{tabular}
\end{table}

Next, we have analyzed the magnetic susceptibility data using CEF schemes. For a tetragonal system, the (2\textit{J}+1) levels of \textit{J} = 5/2 (Ce$^{3+}$) split into three doublets and \textit{J} = 4 (Pr$^{3+}$) multiplet splits into two doublets along with five singlets \cite{CEF_Levels}. The CEF Hamiltonian corresponding to tetragonal site symmetry and $C_{4v}$ point symmetry is as follows
\begin{equation}
\mathcal{H}_{CEF} = B_2^0O_2^0+B_4^0O_4^0+B_4^4O_4^4+B_6^0O_6^0+B_6^4O_6^4
\label{CEF}
\end{equation}

\noindent where $B_p^q$ and $O_p^q$ are the CEF parameters and the Stevens operators, respectively \cite{Stevens_1952,PT_CEF}. The magnetic susceptibility based on the CEF scheme, $\chi_{CEF}^i$ ($i$ = $x$, $y$, $z$), is given by

\begin{equation}
\begin{split}
\chi_{CEF}^i
= &\frac{N_A(g_J\mu_B)^2}{Z} \bigg[ \sum_{n}\beta|\bra{n} J_i \ket{n} |^2 e^{-\beta E_n}  \\
&+\sum_{n\ne m}|\bra{m} J_i \ket{n}|^2 \frac{e^{-\beta E_n}-e^{-\beta E_m}}{E_m-E_n} \bigg],
\end{split}
\label{MT_CEF_Eq}
\end{equation} 

\noindent where $g_J$ is the Land\'{e} factor, $Z=\sum_{n}e^{-\beta E_n}$, and $\beta$ = 1/$k_BT$. $\ket{n}$ is the $n$th eigenfunction and $E_n$ is the corresponding eigenvalue. $J_i$ is the component of angular momentum \cite{CeIr3Ge7}. To analyze the CEF, we have plotted inverse susceptibility as a function of temperature, subtracting temperature-independent term ($\chi_0$) as shown in Figs. \ref{MT}(c) and \ref{MT}(f). The magnetic susceptibility ($\chi_i$ -$\chi_{0}^i$), including the molecular field contribution $\lambda_i$ and the CEF contribution, can be expressed as

\begin{equation}
\label{CEF_MS}
(\chi_i-\chi_{0}^i)^{-1} = (\chi_{CEF}^i)^{-1}-\lambda_i
\end{equation}

\noindent The calculated temperature-dependent magnetic susceptibilities along different crystallographic directions using the above equation, presented in Figs. \ref{MT}(c) and \ref{MT}(f), agree well with the experimental data. The obtained values of CEF parameters and the energy levels of \mbox{CeGaGe} and PrGaGe are given in Tables \textcolor{blue}{III} and \textcolor{blue}{IV}, respectively. The estimated parameters and energy levels are comparable to those of other Ce- and Pr-based compounds \cite{RRhIn5,CePrPd5Al2}. We have also recalculated the CEF parameter $B_2^0$ directly from the paramagnetic Curie-Weiss temperatures to check the consistency of our analysis using following expression \cite{WANG1971_CEF},

\begin{equation}
\Theta_P^c-\Theta_P^{ab} = \frac{3}{10}B_2^0(2J-1) (2J+3)
\label{Theta}
\end{equation}

\noindent The calculated value of $B_2^0$ is -7.3 and -2.4 K for \mbox{CeGaGe} and \mbox{PrGaGe}, respectively, which are in good agreement with estimated value from the CEF model fit.

\begin{table*}
	\label{CEF_PrGaGe}
	\centering 
	\caption {CEF parameters, energy levels and the corresponding wave functions for PrGaGe}
	\vskip .1cm
\begin{tabular}{c c c  c c c }
		\hline
		\hline \\[0.01ex]
		\multicolumn{6}{c}{CEF parameters}\\[1.5ex]
		\hline \\[0.01ex]
		~~~~~$B_2^0$ (K)~~~~~ &~~~~~~ $B_4^0$ (K) ~~~~~&~~~~~$B_4^4$ (K) ~~~~~~&~~~~~~$B_6^0$ (K)~~~~~&~~~~~~$B_6^4$ (K)~~~~~&~~~~~~$\lambda_i$ (mol/emu)~~~~ \\[1.5ex]
		-2.752 $\pm$ 0.338 & 0.050 $\pm$ 0.012 & 0.106 $\pm$ 0.019 & 0 $\pm$ 0.0002& 0.006 $\pm$ 0.0018~~ &$\lambda_z$ = 8.9 $\pm$ 2.1; $\lambda_{x,y}$ = 7.5 $\pm$ 1.9        \\[1.5ex]
		
		\hline
		\hline\\
	\end{tabular}
	
\begin{tabular}{c c c c  c c c c cc}
\multicolumn{10}{c}{Energy levels and wave functions} \\[1.5ex]
\hline \\[0.01ex]
~~~E (K)~~~~~~&~~~~$\ket{+4}$~~~~&~~~~$\ket{+3}$~~~~&~~~~$\ket{+2}$~~~&~~~~$\ket{+1}$~~~~&~~~~
$\ket{0}$~~~~&~~~~$\ket{-1}$~~~~&~~~~$\ket{-2}$~~~~&~~~~$\ket{-3}$~~~~&~~~~$\ket{-4}$ \\[1.5ex]
		\hline\\
202.8 $\pm$ 29.2  &   -0.177   &   0  &     0       &   0  &-0.968&   0  &       0      &  0   & -0.177 \\[1.5ex]
158.9 $\pm$ 23.0 &     0      &   0  &     0       & 0.996&   0  &   0  &       0      & 0.089&   0    \\[1.5ex]
158.9 $\pm$ 23.0  &     0      & 0.089&     0       &   0  &   0  & 0.996&       0      &  0   &   0    \\[1.5ex]
 77.8 $\pm$ 10.5&     0      &   0  &-1/$\sqrt{2}$&   0  &   0  &   0  &-1/$\sqrt{2}$ &  0   &   0    \\[1.5ex]
 68.9 $\pm$ 10.3 &     0      &   0  &-1/$\sqrt{2}$&   0  &   0  &   0  & 1/$\sqrt{2}$ &  0   &   0     \\[1.5ex]
  48.9 $\pm$ 26.1&-1/$\sqrt{2}$&   0  &     0       &   0  &   0  &   0  &       0      &  0   &1/$\sqrt{2}$\\[1.5ex]
 38.6 $\pm$ 26.2 &   0.685    &   0  &     0       &   0  &-0.250&   0  &       0      &  0   & 0.685   \\[1.5ex]
    0 &     0      &   0  &     0       & 0.089&   0  &   0  &       0      &-0.996&   0     \\[1.5ex]
	0 &     0      &-0.996&     0       &   0  &   0  & 0.089&       0      &  0   &   0     \\[1.5ex]
		
		\hline
		\hline
		
	\end{tabular}
\end{table*}

\begin{figure}
	\includegraphics[width=7.2cm, keepaspectratio]{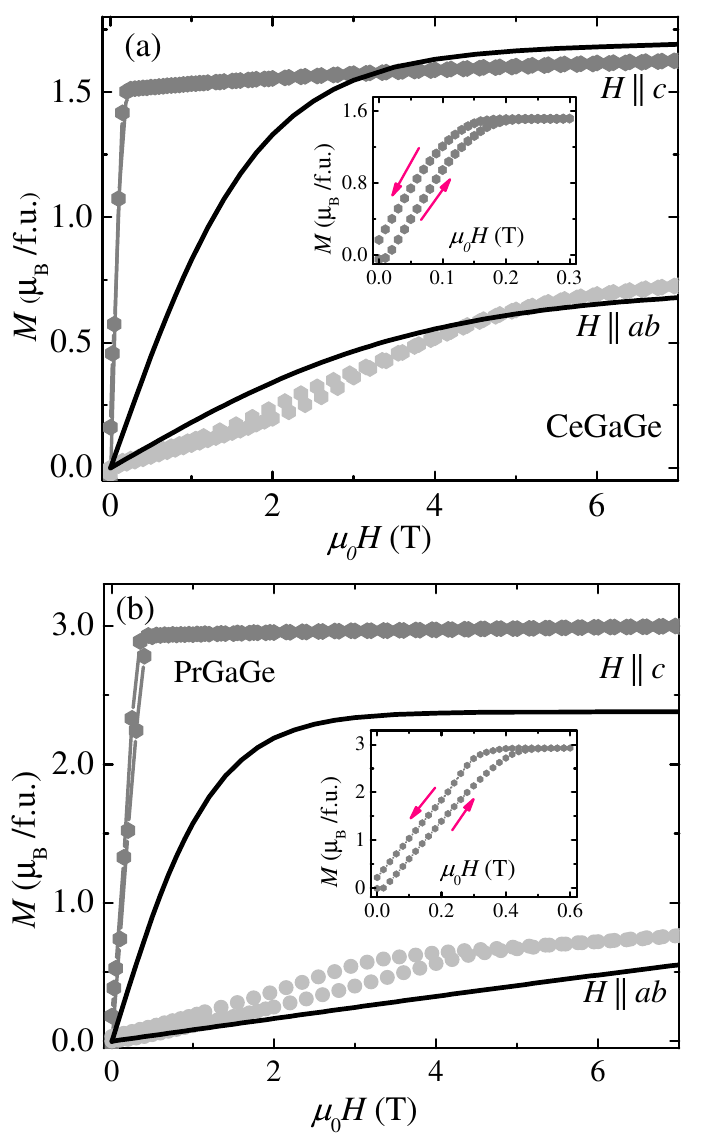}
	\caption{\label{MH}The isothermal magnetization of (a) CeGaGe and (b) PrGaGe was measured at 1.7 K as a function of the applied magnetic fields. Insets of (a) and (b) show a zoomed view of low fields magnetization for the $H \parallel c$ of CeGaGe and PrGaGe, respectively. The black lines are the result of the CEF calculations.}
\end{figure}

Furthermore, we measured the isothermal magnetizations $M(H)$ of CeGaGe and PrGaGe, as shown in Figs. \ref{MH}(a) and \ref{MH}(b), respectively. Magnetization along $H\parallel c$ increases sharply and saturates at 0.15 and 0.35 T for CeGaGe and PrGaGe, respectively. The saturated value of magnetization is 1.55 $\mu_B$ for CeGaGe and 2.95 $\mu_B$ for PrGaGe, which is less than the value of free Ce$^{3+}$ ($gJ\mu_B$ = 2.16 $\mu_B$) and Pr$^{3+}$ ($gJ\mu_B$ = 3.20 $\mu_B$) ions. Such low value of saturation magnetization may be due to the presence of CEF effect \cite{CeAlSi2021,PrAlGe2019,PrAlGe2020}. Whereas, magnetization along $H\parallel ab$ is substantially lower and does not saturate even at 7 T. Such a variation in magnetization in different directions indicates strong magnetocrystalline anisotropy. Moreover, $M(H)$ data also indicate that the \textit{c} axis is the magnetic easy axis as magnetization saturates easily at low applied field. A small hysteresis is observed in both measured directions in the magnetization data, supporting the FM nature in CeGaGe and PrGaGe. We have also determined the isothermal magnetization considering the CEF scheme using the expression
 
\begin{equation}
M_i= \frac{g_J\mu_B}{Z} \sum_{n}|\bra{n} J_i \ket{n} | e^{-\beta E_n}  \\
\label{Mag}
\end{equation} 

\noindent The eigenvalue and associated eigenfunction of the above expression can be calculated by diagonalizing the given total Hamiltonian

\begin{equation}
 \mathcal{H} = \mathcal{H}_{CEF}-g_j \mu_BJ_i(H+\lambda_i M_i),
 \label{M_CEF}
\end{equation}

\noindent where the first term of the Hamiltonian is defined in Eq. (\ref{CEF}). The second and third terms of the above Hamiltonian are the Zeeman and the molecular field term, respectively. The calculated magnetizations of CeGaGe and PrGaGe are represented in Figs. \ref{MH}(a) and \ref{MH}(b) by solid black lines obtained by solving Eqs. (\ref{Mag}) and (\ref{M_CEF}). The above calculation roughly reproduces the magnetization as we neglected the exchange interaction in the Hamiltonian. However, this calculation perfectly recognizes the easy and hard axes of magnetization. Furthermore, we have estimated the saturation magnetization ($M^{sat}$) using the Zeeman term \cite{CeIr3Ge7,CeTiGe3}
\begin{equation}
	\label{Zeeman}
	\mathcal{H}_{x,z} = \textit{g}_{J}\mu_B\bra{\Gamma_{mix,1}}\textit{J}_{x,z}\ket{\Gamma_{mix,1}}\textit{B}_z
\end{equation}
where $\Gamma_{mix,1}$ is the ground state wave function. The estimated and observed value of saturation magnetization is presented in Table \textcolor{blue}{V}. A slight discrepancy between the computed and observed numbers is due to the use of the simplified CEF model \cite{Ce2Re3Si5}.
\begin{table}
	\label{Saturation_M}
	\caption{The calculated saturation magnetizations using Eq. (\ref{Zeeman}) and observed value at 1.7 K and 7 T.} 
	\vskip .1cm
	\begin{tabular}{ccccc}
		\hline
		\hline\\[0.1ex]
		&\multicolumn{2}{c}{CeGaGe}&	\multicolumn{2}{c}{PrGaGe}\\
		
		&Estimated&Observed~~&~~Estimated& Observed\\[1.5ex]
		\hline\\[0.01ex]

		$M_{\parallel c}^{sat} (\mu_B)$        &1.64  &1.63   &2.38   &2.96\\[1.5ex]
		$M_{\perp c}^{sat} (\mu_B)$            &0.68  &0.72   &0 	  &0.74\\[1.5ex]
		
		\hline
		\hline	
	\end{tabular}
\end{table}

\subsection{Specific heat}
\begin{figure*}
	\includegraphics[width=17.5cm, keepaspectratio]{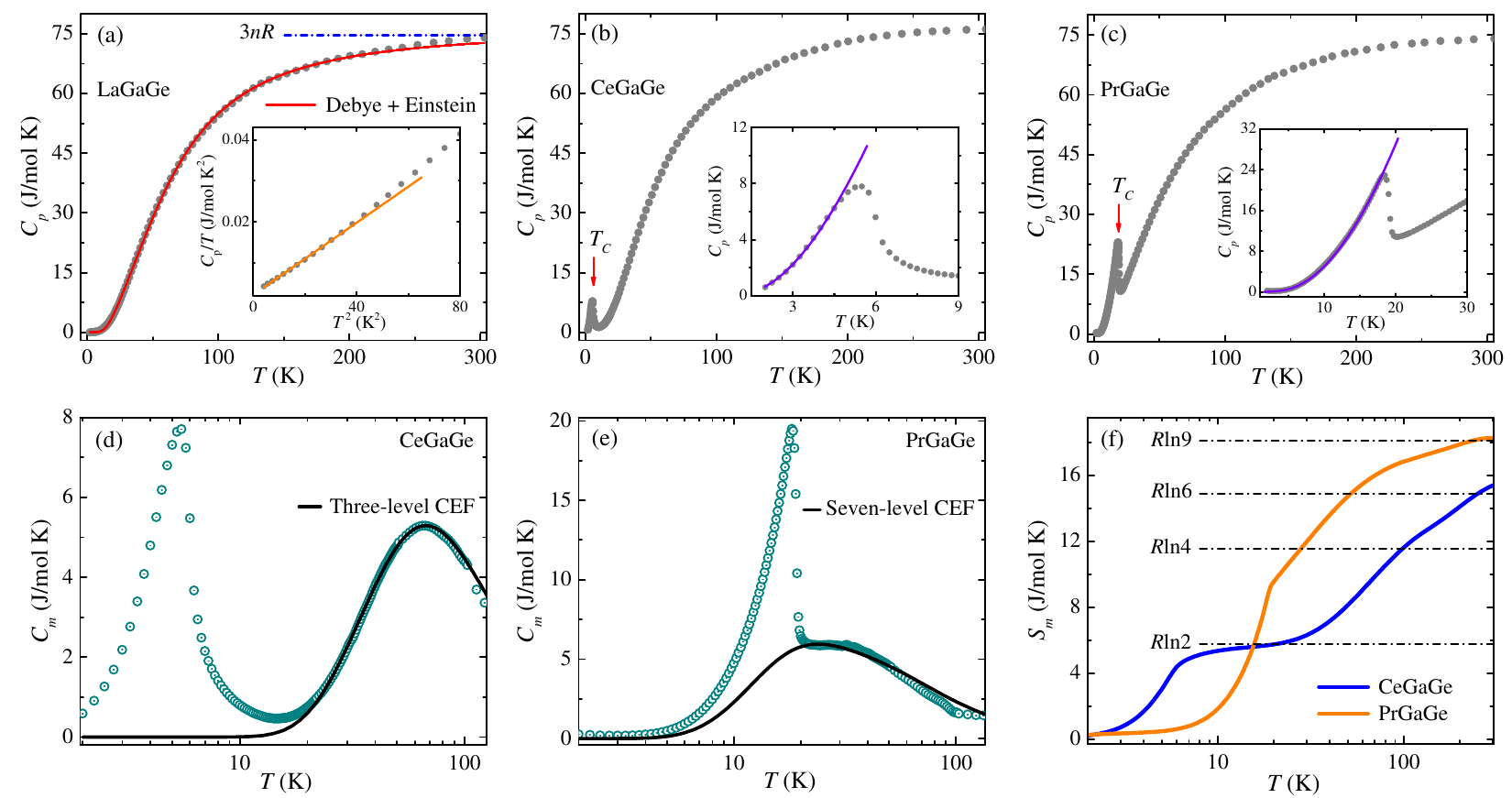}
	\caption{\label{HC}The temperature-dependent heat capacity of single crystals (a) LaGaGe, (b) CeGaGe, and (c) PrGaGe. The red solid line in (a) represents the fitting of the Debye plus Einstein model. The inset of (a) shows linear fitting to $C_p/T$ vs $T^2$ at low temperatures. The purple solid line in the insets of (b) and (c) is the fitting of Eq. \ref{Gap_magnon} to the \textit{C}$_p$  data below \textit{T}$_C$. The temperature dependence of magnetic contribution to \textit{C}$_p$ of (d) CeGaGe and (e) PrGaGe. The solid black line shows the fitting of $C_m(T)$ data using different levels CEF scheme. (f) The magnetic entropy of CeGaGe and PrGaGe calculated as a function of temperature.}
\end{figure*}

The temperature-dependent heat capacity (\textit{C}$_p$) of single crystalline \mbox{RGaGe} (R = La$-$Pr) at a constant pressure is presented in Figs. \ref{HC}(a)$-$\ref{HC}(c), respectively. A sharp anomaly at 6.0 and 19.4 K for CeGaGe and PrGaGe, respectively in the \textit{C}$_p$ data confirms the bulk magnetic ordering as observed in magnetic susceptibility data. The heat capacity reaches an expected value of Dulong-Petit limits $C_p$ = 3\textit{nR} = 74.83 J/mol K for all three compounds, where \textit{n} is the number of atoms in formula unit and \textit{R} is the universal gas constant. The \textit{C}$_p$(\textit{T}) of the nonmagnetic \mbox{LaGaGe} can be expressed well by considering the Debye ($C_D$) and Einstein ($C_E$) modes of heat capacity as presented by the formula

\begin{equation}
C_p(T) = \gamma T+mC_{D}(T)+(1-m)C_{E}(T)
\label{Eq1}
\end{equation}
\noindent where $m$ is the weight factor and $\gamma$ is the Sommerfeld coefficient. $C_D$(\textit{T}) and $C_E$(\textit{T}) are defined as

\begin{equation}
C_{D}(T)=9nR\left( \frac{T}{\Theta_D}\right)^3\int_{0}^{\Theta_D/T}\frac{x^4e^x}{(e^x-1)^2}dx,
\end{equation}
and
\begin{equation}
C_{E}(T)=3nR\left( \frac{\Theta_E}{T}\right)^2\frac{e^{\Theta_E/T}}{(e^{\Theta_E/T}-1)^2},
\end{equation} 

\noindent where $\Theta_D$ and $\Theta_E$ are the Debye and Einstein temperatures, respectively \cite{EuAuAs,GdAgGe}. The obtained fitting parameters are $\Theta_D$ = 295 K, $\Theta_E$ = 100 K and $m$ = 0.76. We have estimated the $\gamma$ from \textit{C}$_p$ data of LaGaGe by applying the formula $C_p/T=\gamma + \beta T^2$ in low-temperature regime (2 K $\leq T \leq$ 7 K). The linear fit of $C_p/T$ vs $T^2$ curve as shown in the inset of Fig. \ref{HC}(a) gives $\gamma$ = 2(2) mJ/mol K$^2$  and $\beta$ = 0.44(1) mJ/mol K$^4$. The estimated value $\gamma$ is comparable to the previous report \cite{CeGaGe1993}. Furthermore, to estimate $\gamma$ for CeGaGe and PrGaGe, we have fitted \textit{C}$_p$ data in the magnetically ordered state as presented in the inset of Figs. \ref{HC}(b) and \ref{HC}(c) using the following formula 

\begin{equation}
	C_p(T) = \gamma T + \beta T^3 + \delta T^{3/2} e^{-\Delta/T}.
	\label{Gap_magnon}
\end{equation}

\noindent The last term of the above equation is the contribution to the \textit{C}$_p$ due to the spin-wave for a ferromagnet with an energy gap $\Delta$ in the magnon spectrum  \cite{gopal_specific}. To perform the above fitting we have used same $\beta$ obtained from \mbox{LaGaGe}, which reduces the fitting parameter. So-obtained fitting parameters are listed in Table \textcolor{blue}{VI}. The estimated value of $\gamma$ for CeGaGe and PrGaGe are comparable to other FM compounds such as CeIr$_2$B$_2$ \cite{CeIr2B2}, Pr$_2$Rh$_2$Ga \cite{Pr2Rh2Ga} and Ce$_{11}$Pd$_{4}$In$_{9}$ \cite{Ce11Pd4In9}. However, it is less than typical Kondo/heavy fermion compounds like CeIn$_3$ \cite{CeIn3} and PrV$_2$Al$_{20}$ \cite{PrV2Al8}.

The magnetic contribution to the \textit{C}$_p$ for CeGaGe and PrGaGe was calculated by subtracting the \textit{C}$_p$ of the nonmagnetic reference LaGaGe, assuming that the phonon contribution in CeGaGe and PrGaGe is approximately equal to that of LaGaGe. The temperature dependence of magnetic heat capacity [$C_m(T)$] is presented in Figs. \ref{HC}(d) and \ref{HC}(e). In addition to the prominent peak at low temperatures due to the magnetic phase transition, a broad hump around 60 and 30 K for CeGaGe and PrGaGe is seen in $C_m(T)$, respectively. The broad peak can be attributed to a Schottky-type anomaly, which arises from the CEF-induced splitting of energy levels Ce$^{3+}$ and Pr$^{3+}$. The generalized formula of Schottky contribution to the heat capacity for multi-level CEF scheme is given by

\begin{equation}
	\begin{split}
		C_{Sch}(T)
		=&\left( \dfrac{R}{T^2}\right) \bigg[  \sum_{i} g_ie^{-\Delta_i/T}\sum_{i}g_i{\Delta_i}^2e^{-\Delta_i/T}\\
		& - \left( \sum_{i}g_i\Delta_ie^{-\Delta_i/T}\right)^2\bigg]  \left( \sum_{i}g_ie^{-\Delta_i/T}\right)^{-2},
	\end{split}
\end{equation} 

\noindent where $g_i$ is the degeneracy of the \textit{i}th state with $\Delta_i$ energy gap splitting \cite{gopal_specific}. Here, we have used the same CEF parameters obtained from CEF analysis of magnetic susceptibility data to reproduce the Schottky anomaly observed in $C_m$(\textit{T}). Interestingly, the calculated Schottky anomaly matches well with the experimental data for both compounds, as presented in Figs. \ref{HC}(d) and \ref{HC}(e). This justifies the validity of the CEF parameters and the energy level splittings of both compounds.

\begin{figure*}
	\centering
	\includegraphics[width=17.0cm, keepaspectratio]{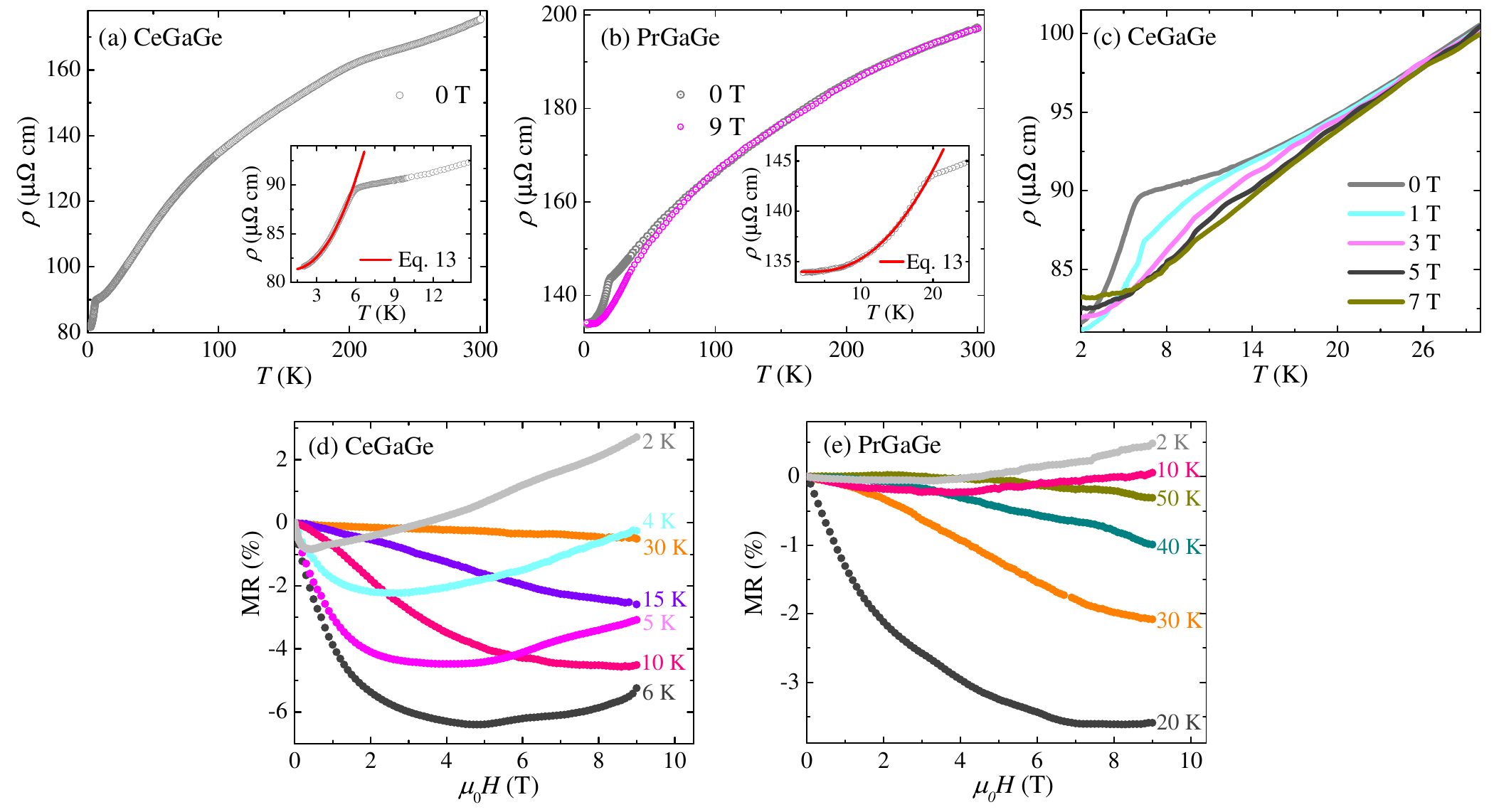}
	\caption{\label{RT}Temperature-dependent electrical resistivity of (a) CeGaGe and (b) PrGaGe measured in between 2$-$300 K. The insets show the fitting of Eq. \ref{RT_fit} in the magnetically ordered state. (c) The temperature dependence of electrical resistivity of CeGaGe measured under external magnetic fields. The MR of (d) CeGaGe and (e) PrGaGe as a function of the magnetic field for different temperatures.}
\end{figure*}

\begin{table}
	\label{HC_para}
	\centering 
	\caption {Obtained fitting parameters of low temperature \textit{C}$_p$ data of CeGaGe and PrGaGe using Eq. (\ref{Gap_magnon}).}
	\vskip .1cm
	\begin{tabular}{c c c  c  }
		\hline
		\hline \\[0.01ex]
		&$\gamma$ (mJ/mol K$^2$) ~~~& $\delta~(J/mol K^{5/2})$ ~~~& $\Delta$ (K) \\[1.5ex]
		\hline \\[0.01ex]
		
		CeGaGe   &13 &1.54  &3.90 \\[1.5ex]
		PrGaGe   &35 &0.58  &14.42 \\[1.5ex]

		\hline
		\hline
		
	\end{tabular}
\end{table}
Next, we have calculated magnetic entropy using the expression $S_{m}(T) =\int\frac{C_{m}}{T} dT$. The magnetic entropy as a function of temperature for CeGaGe and PrGaGe is presented in Fig. \ref{HC}(f). According to CEF analysis, six energy levels in CeGaGe split into three doublets, resulting in three plateaus in $S_{m}$(\textit{T}) near \textit{R}ln2, \textit{R}ln4, and \textit{R}ln6. In case of PrGaGe, the $S_{m}$(\textit{T}) reaches a value of 7.9 J/mol K at $T_C$, which lies between a value of magnetic entropy of ground state \textit{R}ln2 and \textit{R}ln4. Therefore, a strong thermal population of the excited levels is indicated, and the first excited level is probably slightly higher in energy than $T_C$.  The total magnetic entropy \textit{R}ln6 for CeGaGe and \textit{R}ln9 for PrGaGe is released at 252 and 210 K, respectively, indicating the overall splitting of CEF levels within this energy window. The observed $S_{m}$(\textit{T}) is consistent with our CEF analysis.

\subsection{Magnetotransport}

The electrical resistivity $\rho(T)$ measured within the \textit{ab} plane in CaGaGe and PrGaGe single crystals in the temperature range of 2$-$300 K is shown in Figs. \ref{RT}(a) and \ref{RT}(b), respectively. The $\rho(T)$ of CeGaGe and PrGaGe show metallic behavior down to 2 K with a sharp anomaly around 6.0 and 19.4 K, respectively. The low-temperature anomaly is associated with the magnetic phase transition where resistivity drops rapidly due to reduction of spin scattering as systems order magnetically. The residual resistivity ratio (RRR) of CeGaGe and PrGaGe is $\sim$ 2.1 and $\sim$ 1.5, respectively, which is small compared to typical metals. Such value of RRR may appear due to site disorder between Ga and Ge. Similar RRR has also been observed in RAlX (R = La$-$Nd and X = Ge, Si) family for site disorder between Al and Ge/Si \cite{CeAlGe2018,CeAlSi2021,PrAlGe2019}. On the other hand, $\rho(T)$ of \mbox{CeGaGe} exhibits a broad hump from 220 to 50 K, may arise from the CEF splitting, which is also evident from thermodynamic measurement. Similarly a concave curvature is observed in the $\rho(T)$ data of PrGaGe. Such behavior of $\rho(T)$ is commonly seen in rare-earth compounds like CeAuGa$_3$ \cite{CeAuGa3}, Ce$_2$NiSi$_3$ \cite{Ce3NiSi3} and Pr$_7$Ru$_3$ \cite{Pr7Ru3}. For both compounds, the $\rho(T)$ in the magnetically order state can be described using the following expression 

\begin{equation}
	\label{RT_fit}
	\rho(T) = \rho_0 + AT^2\text{exp}\left( -\frac{\Delta}{T}\right), 
\end{equation} 

\noindent where $\rho_0$ is the residual resistivity, and the second term is the contribution to resistivity due to FM magnon. $A$ is related to the strength of electron-magnon scattering, and $\Delta$ is the energy gap in the magnon spectrum \cite{TmAuGe}. The so-obtained value of fitting parameters from Eq. \ref{RT_fit} are listed in Table \textcolor{blue}{VII}. The estimated value of $\Delta$ is very close to that obtained from the heat capacity data, confirming consistency of our analysis.
\begin{table}
	\label{RT_para}
	\centering 
	\caption {The estimated fitting parameters $\rho_0$,  $A$ and $\Delta$ from the resistivity data using Eq. (\ref{RT_fit}).}
	\vskip .1cm
	\begin{tabular}{c c c  c  }
		\hline
		\hline \\[0.01ex]
		~~&~~$\rho_0$ ($\mu$$\Omega$ cm) ~~&~~ $A$ ($\mu$$\Omega$ cm/K$^2$) ~~&~~ $\Delta$ (K) ~~\\[1.5ex]
		\hline \\[0.01ex]
		
		CeGaGe   &81.3 &0.45  &3.33 \\[1.5ex]
		PrGaGe   &134.0 &0.05  &13.36 \\[1.5ex]

		\hline
		\hline
		
	\end{tabular}
\end{table}

The $\rho(T)$ of \mbox{CeGaGe} and \mbox{PrGaGe} under various external magnetic fields is presented in Figs. \ref{RT}(b) and \ref{RT}(c). The low temperature anomaly in $\rho$(\textit{T}) due to magnetic ordering for both compounds is getting smeared out by the applied field, and the magnitude of electrical resistivity is reduced near $T_C$. The field dependence of MR of CeGaGe and PrGaGe at various temperatures for $H \parallel c$ are represented in Figs. \ref{RT}(d) and \ref{RT}(e), respectively. The MR is defined as [$\rho(H) - \rho(0)]/\rho(0)$, where $\rho(H)$ and $\rho(0)$ are the resistivities in the presence and absence of a magnetic field, respectively. The MR of both compounds becomes negative close to the magnetic ordering temperature and remains negative throughout the paramagnetic region. The maximum value of MR reaches about -6.5 \%  and -3.8 \% for CeGaGe and PrGaGe, respectively. The negative MR for both samples near $T_C$ results from the suppression of spin scattering by the applied magnetic field, which is the typical MR behavior for an FM \cite{PrRhSn3,RAgAl3}. Interestingly, at low temperatures (T $< T_C$), the MR for both samples in a weak field regime is negative, but as the field intensity increases, the MR rises and becomes positive. The magnetic moments become polarized at low applied fields ($\sim$ 0.4 T), and MR becomes negative due to suppression of spin disorder; however, as the field strength increases, the Lorentz force becomes dominant, and MR starts to increase, as observed in PrAlGe \cite{PrAlGe2019} and CeIr$_2$B$_2$ \cite{CeIr2B2}.

\section{Summary}

We have studied the magnetic, thermodynamic, and magnetotransport properties of RGaGe (R = Ce and Pr) single crystals synthesized using gallium-indium flux. The powder XRD patterns indicated that these crystals crystallize in a body-centered tetragonal structure with space group \textit{I}4$_1$\textit{md} (No. 109). The temperature-dependent magnetic susceptibility and resistivity measurements show a magnetic phase transition in CeGaGe and PrGaGe at $T_C$ = 6.0 and 19.4 K, respectively. Further, heat capacity data shows a sharp peak at the magnetic transition, which confirms the bulk nature of the magnetic ordering. The temperature and field-dependent magnetization data reveal strong magnetic anisotropy and the presence of CEF in both compounds. The CEF analysis for inverse magnetic susceptibility and heat capacity data of CeGaGe and PrGaGe reveal that \textit{J} = 5/2 degenerate ground state of Ce$^{3+}$ ion splits into three doublets, whereas \textit{J} = 4 multiplet of Pr$^{3+}$ ion splits into two doublets and five singlets. The field-dependent MR for both compounds is positive at low temperatures and high fields, it switches to negative near the magnetic transition temperature and remains negative in the paramagnetic region.

\section{Acknowledgment} 

We acknowledge IIT Kanpur and the Department of Science and Technology, India, [Order No. DST/NM/TUE/QM-06/2019 (G)] for financial support. We thank Suman Sanki for valuable discussion. D.K. acknowledges financial support from the National Science Centre (Poland) under Research Grant No. 2021/41/B/ST3/01141.

\bibliographystyle{apsrev4-2}
\bibliography{Reference_RGaGe}

\end{document}